\documentclass[
    ,final           
  ]
  {aipproc}

\layoutstyle{8x11double}

\usepackage{epsfig}

\begin{document}

\title{Precise determination of the $\eta_c$ mass and width in the radiative $J/\psi\to\eta_c\gamma$ decay}


\author{Nora Brambilla}{
  address={Physik-Department, Technische Universit\"at M\"unchen, James-Franck-Str. 1, 85748 Garching,
Germany}
}

\author{Pablo Roig}{address={Laboratoire de Physique Th\'eorique, 
CNRS/Univ. Paris-Sud 11 (UMR 8627), B\^at 210, Facult\'e des Sciences, F-91405 Orsay Cedex}}

\author{Antonio Vairo}{
  address={Physik-Department, Technische Universit\"at M\"unchen, James-Franck-Str. 1, 85748 Garching,
Germany}
}

\begin{abstract}
We present an effective field theory based extraction of the $\eta_c$ mass and width
from a recent measurement by CLEO of the photon line shape in the $J/\psi\to\eta_c\gamma$ decay.
\end{abstract}

\maketitle

We study the radiative decay $J/\psi\to\eta_c \gamma$ within
an effective field theory (EFT) framework, namely potential
non-relativistic QCD (pNRQCD) \cite{Brambilla:2004jw, Pineda:1997bj, Brambilla:1999xf}. 
Our motivation is to take advantage of the new measurements made by the CLEO \cite{:2008fb} 
and KEDR \cite{Anashin:2010nr} collaborations of the photon line shape, 
in order to obtain precise determinations of the $\eta_c$ mass and width 
(for a recent review see \cite{Brambilla:2010cs}).
Magnetic dipole transitions in quarkonium were studied in pNRQCD 
in Ref.~\cite{Brambilla:2005zw}, where, for the $J/\psi\to\eta_c\gamma$ decay, 
a branching ratio consistent, within errors, with the PDG value \cite{Nakamura:2010zzi} 
was found. We apply the same formalism here. Before discussing it, however,  
we briefly review the CLEO and KEDR analyses.
 
To fit its data \cite{:2008fb}, CLEO uses two background sources,
\begin{itemize}
\item a Monte Carlo modeled background for spurious $J/\psi\to X$ 
with shape 
$$
\mathrm{bkg}_1(E_\gamma) = N \left( e^{-5.720\,E_\gamma} + 10.441\,e^{-33.567\,E_\gamma}\right) 
$$ 
\item and a freely fit background for $J/\psi\to\pi^0X$ and non-signal 
$J/\psi\to X$ with shape  
$$
\mathrm{bkg}_2(E_\gamma)= A + B E_\gamma + C E_\gamma^2,
$$
\end{itemize}
and a theoretical line shape given by
$$
\mathrm{theory}(E_\gamma) = E_\gamma^3 \times {\rm BW}_{\mathrm{rel}}(E_\gamma) \times \mathrm{damping}(E_\gamma),
$$ 
where the relativistic Breit--Wigner distribution is 
\begin{eqnarray*}
{\rm BW}_{\mathrm{rel}}(E_\gamma)^{-1} &=& 
\left(M_{J/\psi}^2\,-\,2\,M_{J/\psi}\,E_\gamma\,-\,M_{\eta_c}^2\right)^2\,
\\
&& +\,\left(M_{J/\psi}^2\,-\,2\,M_{J/\psi}\,E_\gamma\right)\,\Gamma_{\eta_c}^2, 
\end{eqnarray*}
$M_{J/\psi}$ and $M_{\eta_c}$ stand for the masses of the $J/\psi$ and 
$\eta_c$ respectively, $\Gamma_{\eta_c}$ for the $\eta_c$ width, and 
$$
\mathrm{damping}(E_\gamma)\,=\,e^{-E_\gamma^2/(8\,\beta^2)}.
$$
The damping function accounts for the overlap of the two 
quarkonium states, assumed to be described by wavefunctions of 
an harmonic oscillator, and the photon. 
The natural scale of $\beta$ is then the typical momentum transfer 
inside the charmonium ground state, which is about $700$ to $800$ MeV. 
This implies almost no damping for photon energies smaller than $500$ MeV, which 
is consistent with the multipole expansion of the electromagnetic fields.
The value that comes from CLEO's fit is, however, an order of magnitude smaller, 
$\beta = (65.0\pm2.5)$ MeV, which implies the vanishing of the signal 
for photon energies of few hundred MeV. The CLEO analysis yields the values 
$M_{\eta_c}\,=\,(2982.2\pm0.6)$ MeV and $\Gamma_{\eta_c}\,=\,(31.5\pm1.5)$ MeV. 

The $\eta_c$ line shape is also studied by the KEDR collaboration
in \cite{Anashin:2010nr}. Their analysis is similar to 
CLEO's one, but an alternative damping function is used as well:
$$
\mathrm{damping'}(E_\gamma) = \frac{E_{\rm peak}^2}{E_\gamma\,E_{\rm peak}\,+\,(E_\gamma\,-\,E_{\rm peak})^2},
$$
where $E_{\rm peak}$ is the most probable transition energy. 
If CLEO's data are used, the analysis of KEDR gives 
$M_{\eta_c}\,=\,(2982.4\pm0.7)$ MeV and $\Gamma_{\eta_c}\,=\,(32.5\pm1.8)$ MeV 
when fitting with $\mathrm{damping}(E_\gamma)$, and 
$M_{\eta_c}\,=\,(2981.8\pm0.5)$ MeV and $\Gamma_{\eta_c}\,=\,(33.6\pm1.9)$ MeV   
when fitting with $\mathrm{damping'}(E_\gamma)$.
However, an analysis of KEDR's own preliminary data gives different values:
$M_{\eta_c}\,=\,(2979.7\pm1.6)$ MeV and $\Gamma_{\eta_c}\,=\,(26.9\pm4.8)$ MeV 
when fitting with $\mathrm{damping}(E_\gamma)$, and $M_{\eta_c}\,=\,(2979.4\pm1.5)$ MeV 
and $\Gamma_{\eta_c}\,=\,(27.8\pm5.1)$ MeV when fitting with $\mathrm{damping'}(E_\gamma)$.\footnote{
Incidentally, the KEDR results are closer to previous $\eta_c$ mass measurements 
from $J/\psi$ and $\psi(2S)\to\eta_c\gamma$ decays (averaging $(2977.3\pm1.3)$
MeV \cite{Amsler:2008zzb}), while the CLEO value is in agreement with
the results obtained in $\gamma\gamma$ fusion and
$p\bar{p}$ production (averaging $(2982.6\pm1.0)$ MeV \cite{Amsler:2008zzb}).}  
The discrepancy between different values of the $\eta_c$ mass and width 
is larger than the experimental sensitivity, which highlights the importance 
of performing a critical analysis of the theory inputs. 

Potential NRQCD exploits the hierarchy of scales in the problem and allows 
to express physical observables as systematic expansions in the ratio 
of these scales. 
Heavy quark-antiquark bound states are characterized by a number of scales:
the heavy-quark mass, the typical momentum, $\langle p \rangle$, exchanged by the 
quarks (this is also of the order of the inverse of the typical 
size of the bound state, $1/\langle r \rangle$), the binding energy,  
the typical hadronic scale $\Lambda_{\rm QCD}$, and 
possibly other smaller scales. In the transition $J/\psi\to\eta_c \gamma$, 
the relevant scales are the charm mass $m_c$, which is much larger 
than the next-to-largest scale, which is
$\left\langle p\right\rangle \sim 1/\left\langle r\right\rangle \sim 800$ MeV, 
which in turn is larger than $\Lambda_{\rm QCD}$. 
The binding energy of the $J/\psi$ is $E_{J/\psi}\sim 500$ MeV, 
which is smaller than $1/\left\langle r\right\rangle$. 
There are also two smaller scales that have to be considered in addition: 
the hyperfine splitting, which is $M_{J/\psi}-M_{\eta_c}\sim 120$ MeV 
and smaller than $E_{J/\psi}$, and the width of the $\eta_c$, 
$\Gamma_{\eta_c}\sim 30$ MeV, which is smaller than the hyperfine splitting. 
In the radiative decay $J/\psi \to X\gamma$ around the $\eta_c$ peak, 
we consider photon energies that vary between $0$ MeV
and $500$ MeV$<1/\left\langle r\right\rangle$.
Under these conditions, we may describe the charmonium ground state 
in weakly coupled pNRQCD (because the system is non-relativistic and the 
typical momentum transfer is larger than $\Lambda_{\rm QCD}$), 
couple the electromagnetic fields to pNRQCD and multipole expand the electromagnetic 
fields (because the photon energy is smaller than the typical momentum transfer in the 
bound state). 

Three main processes contribute to $J/\psi\to X \gamma$ 
within pNRQCD in the energy range of interest ($0$ MeV$\le E_\gamma \le 500$ MeV):
\begin{itemize}
 \item the M1 transition $J/\psi\to\eta_c \gamma\to X \gamma$, 
 \item the E1 transitions $J/\psi\to\chi_{c(0,2)}(1P) \gamma\to X \gamma$, 
 \item fragmentation and other background processes included in the background functions.
\end{itemize}

\begin{figure}[ht]
\makebox[-3truecm]{\phantom b}
\put(0,0){\epsfxsize=6truecm \epsfbox{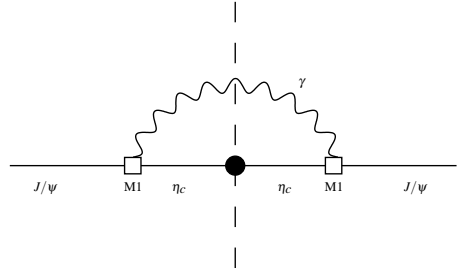}}
\put(10,30){\tiny $J/\psi$}
\put(44,30){\tiny M1}
\put(62,30){\tiny $\eta_c$}
\put(102,30){\tiny $\eta_c$}
\put(110,70){\tiny $\gamma$}
\put(120,30){\tiny M1}
\put(150,30){\tiny $J/\psi$}
\caption{M1 contribution to the $\eta_c$ line shape. The black dot stands 
for four-fermion operators that contribute to the decay width of the intermediate state.}
\label{diag1}
\end{figure}

\begin{figure}[ht]
\makebox[-3truecm]{\phantom b}
\put(0,0){\epsfxsize=6truecm \epsfbox{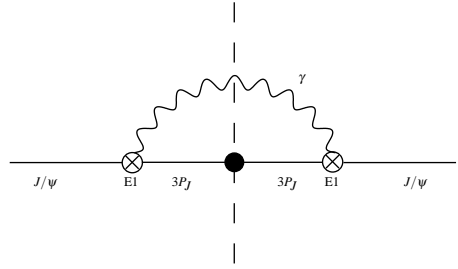}}
\put(10,30){\tiny $J/\psi$}
\put(44,30){\tiny E1}
\put(62,30){\tiny $3P_J$}
\put(102,30){\tiny $3P_J$}
\put(110,70){\tiny $\gamma$}
\put(120,30){\tiny E1}
\put(150,30){\tiny $J/\psi$}
\caption{E1 contribution to the $\eta_c$ line shape.}
\label{diag2}
\end{figure}

The evaluation of the magnetic dipole contribution yields (see Fig.~\ref{diag1})
\begin{equation}
\label{M1}
\frac{\mathrm{d\Gamma^{M1}_{J/\psi\to\eta_c\gamma}}}{\mathrm{d}E_\gamma}
=\frac{64}{27}\frac{\alpha}{\pi}\frac{E_\gamma^3}{M_{J/\psi}^2}
\frac{{\Gamma_{\eta_c}}/{2}}{\left(
M_{J/\psi}-M_{\eta_c}-E_\gamma\right)^2+\frac{\Gamma_{\eta_c}^2}{4}}.
\end{equation} 
It has been pointed out in Ref.~\cite{:2008fb} that the dependence on $E_\gamma^3$ 
is responsible for the asymmetric shape of the photon spectrum.

The electric dipole contribution is (see Fig.~\ref{diag2})
\begin{eqnarray}
\label{E1}
\frac{\mathrm{d\Gamma^{E1}_{J/\psi\to\eta_c\gamma}}}{\mathrm{d}E_\gamma}=
\frac{448}{243} \alpha\frac{E_\gamma}{m_c}
\alpha_{\rm s}^2\frac{\bigg|\phi_{J/\psi}(0)\bigg|^2}{m_c^3}
\bigg|a_e(E_\gamma)\bigg|^2,
\end{eqnarray} 
where $\phi_{J/\psi}(0)$ is the $J/\psi$ wavefunction at the origin.
The function $a_e(E_\gamma)$ has been discussed in
Refs.~\cite{Manohar:2003xv, RuizFemenia:2008zz}; 
a closed analytical form has been derived in \cite{Voloshin:2003hh}.

In the weak-coupling regime, the typical momentum transfer in the charmonium 
is of order $m_c\alpha_{\rm s}$, the binding energy of the ground state 
is of order $m_c\alpha_{\rm s}^2$ and the hyperfine splitting is 
of order $m_c\alpha_{\rm s}^4$. The magnetic and electric dipole contributions are of equal order for
$m_c\,\alpha_{\rm s} \gg E_\gamma \gg m_c\,\alpha_{\rm s}^2$. 
The magnetic contribution completely dominates the electric one in the
peak region ($m_c\,\alpha_{\rm s}^2\gg E_\gamma \gg m_c\,\alpha_{\rm s}^4$) 
and it also dominates by a factor $E_{J/\psi}^2/\left(M_{J/\psi}-M_{\eta_c}\right)^2\sim
1/\alpha_{\rm s}^4$ for $E_\gamma \ll m_c\,\alpha_{\rm s}^4$.

\begin{figure}[ht]
\vspace{0.8cm}
 \centering
  \includegraphics[scale=0.3, angle=-90]{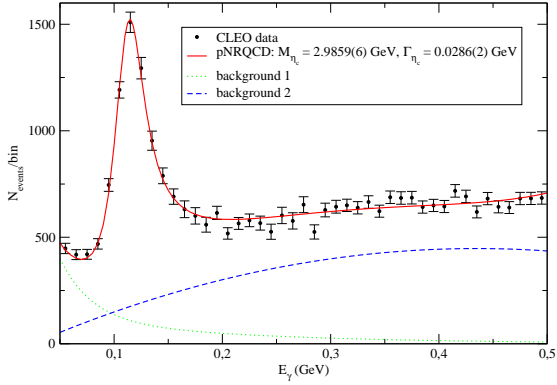}
\caption{Our best fit to CLEO's data for the photon spectrum in $J/\psi\to\eta_c \gamma$ 
using Eqs. (\ref{M1}) and (\ref{E1}) for the theoretical signal together with the two background 
sources $\mathrm{bkg}_1(E_\gamma)$ and $\mathrm{bkg}_2(E_\gamma)$.}
 \label{bestfit}
\end{figure}

We use as theoretical line shape the sum of Eqs. (\ref{M1}) and (\ref{E1}), 
and as background the sum of $\mathrm{bkg}_1(E_\gamma)$ and $\mathrm{bkg}_2(E_\gamma)$.
As predicted by the power counting and numerically confirmed, 
for photon energies $\le 500$ MeV, the electric dipole contribution is negligible 
with respect to the magnetic one.
The signal shape has been convolved with a Gaussian resolution function, whose resolution width 
is 4.8 MeV \cite{:2008fb}. Our best fit is shown in Fig. ~\ref{bestfit}.
The fitting parameters are $M_{\eta_c}$, $\Gamma_{\eta_c}$, an overall normalization, 
the signal normalization, and the background parameters $A$, $B$ and $C$. 
In particular, the best fit line shape parameters for the $\eta_c$ are (errors 
are only statistical)
\begin{eqnarray*}
M_{\eta_c} &=& (2985.9\pm0.6) \; {\rm MeV},  
\\ 
\Gamma_{\eta_c} &=& (28.6\pm0.2) \; {\rm MeV}.
\end{eqnarray*}
The main differences between our analysis and the one of CLEO 
are summarized in the following.
\begin{itemize}
 \item At leading order in the multipole expansion of the photon field, 
which is justified in the range $E_\gamma \le 500$ MeV, 
and in the non-relativistic limit,
the overlap integral in the magnetic dipole transition is equal to one.
This amounts to setting equal to one the damping function in CLEO's analysis.
We have checked that higher-order terms in the multipole expansion are indeed 
negligible. Our conclusion is that CLEO's damping function has no theoretical 
justification as an overlap integral, which is also signalled by the 
unnatural scale of the parameter $\beta$. The absence of a damping 
function accounts for about $50\%$ of the difference in the $\eta_c$ mass
determination in our analysis and in that one of CLEO.
 \item Another difference is that in our analysis the differential width turns 
out to be proportional to a non-relativistic Breit--Wigner 
distribution  ($\times E_\gamma^3$), see Eq. (\ref{M1}),  
while CLEO uses a relativistic Breit--Wigner distribution. 
The relativistic Breit--Wigner distribution resums 
some classes of relativistic corrections, without including others of the 
same size. Partial resummations could lead to spurious effects, while  
the use of a non-relativistic Breit--Wi\-gner distribution appears to be  
more justified in the framework of a systematic relativistic expansion, like 
the one provided by pNRQCD. This difference accounts for about the remaining 
$50\%$ difference in the determination of the $\eta_c$ mass.
\end{itemize} 

In summary, radiative decays of quarkonia and specifically the transition 
$J/\psi\to\eta_c \gamma$ may be investigated in 
an EFT framework that systematically exploits the hierarchy of scales 
in the system (pNRQCD). The total transition width has been calculated 
in \cite{Brambilla:2005zw} including the next-to-leading order relativistic 
corrections. Within a large theoretical uncertainty that determination 
is in agreement with the experimental value. 
In this work, we consider the photon line shape in the non-relativistic limit.
We obtain a best fit, which is in good agreement with CLEO's experimental 
determination, see Fig. ~\ref{bestfit}. However, theoretical errors have 
not been included so far in our analysis and we expect them to be larger than 
those coming from the fitting accuracy.
For instance, relativistic corrections could impact the $\eta_c$ mass 
by corrections as large as those induced by the difference between the  
relativistic and the non-relativistic Breit--Wi\-gner distribution.
Under investigation is also the extraction of the 
$J/\psi\to\eta_c \gamma$  branching ratio from the photon spectrum.

\begin{theacknowledgments}
We thank R.E.~Mitchell for useful correspondence. The authors
acknowledge financial support from the RTN Flavianet
MRTN-CT-2006-035482 (EU), and N.B. and A.V. from the DFG cluster of
excellence ''Origin and struc\-tu\-re of the universe''
(www.universe-cluster.de). The work of P.R. was partly funded by a
Marie Curie ESR Contract (FLAVIAnet).
\end{theacknowledgments}

\bibliographystyle{aipproc}   

\begin{thebibliography}{9}
\bibitem{Brambilla:2004jw}
  N.~Brambilla, A.~Pineda, J.~Soto and A.~Vairo,
  Rev.\ Mod.\ Phys.\  {\bf 77} (2005) 1423.

\bibitem{Pineda:1997bj}
  A.~Pineda and J.~Soto,
  Nucl.\ Phys.\ Proc.\ Suppl.\  {\bf 64} (1998) 428.

\bibitem{Brambilla:1999xf}
  N.~Brambilla, A.~Pineda, J.~Soto and A.~Vairo,
  Nucl.\ Phys.\  B {\bf 566} (2000) 275.

\bibitem{:2008fb}
  R.~E.~Mitchell {\it et al.}  [CLEO Collaboration],
  Phys.\ Rev.\ Lett.\  {\bf 102} (2009) 011801.

\bibitem{Anashin:2010nr}
  V.~V.~Anashin {\it et al.},
  arXiv:1002.2071 [hep-ex].

\bibitem{Brambilla:2010cs}
  N.~Brambilla {\it et al.},
  arXiv:1010.5827 [hep-ph].

\bibitem{Brambilla:2005zw}
  N.~Brambilla, Y.~Jia and A.~Vairo,
  Phys.\ Rev.\  D {\bf 73} (2006) 054005.

\bibitem{Nakamura:2010zzi}
  K.~Nakamura {\it et al.}  [Particle Data Group],
  J.\ Phys.\ G {\bf 37}, 075021 (2010).

\bibitem{Amsler:2008zzb}
  C.~Amsler {\it et al.}  [Particle Data Group],
  ``Review of particle physics,''
  Phys.\ Lett.\  B {\bf 667} (2008) 1. 

\bibitem{Manohar:2003xv}
  A.~V.~Manohar and P.~Ruiz-Femen\'{\i}a,
  Phys.\ Rev.\  D {\bf 69} (2004) 053003;
 P.~D.~Ruiz-Femen\'{\i}a,
  Nucl.\ Phys.\ Proc.\ Suppl.\  {\bf 152} (2006) 200.

\bibitem{RuizFemenia:2008zz}
  P.~D.~Ruiz-Femen\'{\i}a,
  Nucl.\ Phys.\  B {\bf 788} (2008) 21,
  PoS {\bf EFT09} (2009) 005.

\bibitem{Voloshin:2003hh}
  M.~B.~Voloshin,
  Mod.\ Phys.\ Lett.\  A {\bf 19} (2004) 181.

\end{thebibliography}

\end{document}